\documentstyle[epsf]{article}
\newcommand{\dir}{Figs}
\newcommand{\fig}[4]
{
\begin{center}
\noindent
\unitlength=1mm
\begin{picture}(#2,#3)
\put(0,0){\leavevmode \epsfxsize=#2mm \epsffile{\dir/#1}}
\end{picture}
\noindent
\end{center}
#4
}

\begin{document}
\baselineskip=12pt
\setcounter{page}{1}
%

\newcommand{\CCgeneric}
{
\caption{ Generic phase diagram for fatty acid monolayers 
(according to Ref. [7]).
LE is the liquid expanded phase, CS and L${}_2$'' have
positional order, all other phases are hexatic liquids. 
The phases Ov and L${}_2$' show tilt towards nearest neighbours,
L${}_2$ and L${}_2$'' tilt towards next-nearest neighbours,
and LS, S, CS are untilted. In CS, S, L${}_2$', and L${}_2$'',
the backbones of the hydrocarbon chains are ordered. }
\label{fig:generic}
}

\newcommand{\CCphtheory}
{
\caption{ Phase diagram obtained from self-consistent field
calculations in the plane of chain anisotropy vs. molecular
area. LE denotes liquid expanded phase, LS untilted liquid
condensed phase, and L${}_2$ a tilted condensed phase 
(after Ref. [9]).}
\label{fig:phtheory}
}

\newcommand{\CCphsima}
{
\caption{ Phase diagram obtained from Monte Carlo simulations
in the plane of spreading pressure $\Pi$ 
(in units of $\epsilon/k_B \sigma_T^2$)
vs. temperature $T$ (in units of $\epsilon/k_B$).
LE denotes disordered phase, LC-U untilted
ordered phase, LC-NN ordered phase with tilt towards nearest neighbours,
and LC-NNN ordered phase with tilt towards next-nearest neighbours.
At pressures above $\Pi = 20 \epsilon/\sigma_T^2$, the tilt direction 
is unclear. (After Ref. [20].) }
\label{fig:phsima}
}

\newcommand{\CCrpvf}
{
\caption{ Radial pair correlation function $g(r)$ for the heads (a) 
and the projections of the centres of gravity into the $xy$-plane (b)
vs. $r$ (in units $\sigma_T$) at spreading pressure 
$\Pi = 1 \: \epsilon/\sigma_T^2$ and various temperatures (in units 
$\epsilon/k_B$), as indicated. The LE to LC-NN phase transition takes 
place at temperature $T = 1.4 \epsilon/k_B$. The values
$g(r)$ for the temperature $T=0.1 \epsilon/k_B$ are rescaled by
a factor of 5. (After Ref. [20].)}
\label{fig:rpvf}
}

\newcommand{\CCrxy}
{
\caption{ Collective tilt order parameter $R_{xy}$ vs. pressure
$\Pi$ in units $\epsilon/k_B \sigma_T^2$ at temperature
$T = 0.5 \epsilon/k_B$ for different potential widths.
(units $\epsilon/k_B \sigma_T$) vs. temperature $T$ (units $\epsilon/k_B$).}
\label{fig:rxy_2}
}

\newcommand{\CCrxyh}
{
\caption{ Distribution of the order parameter $R_xy$ at spreading
pressure $\Pi = 15 \epsilon/k_B \sigma_T^2$ and temperature
$T = 1.3 \epsilon/k_B$ }
\label{fig:rxy_hist}
}

\newcommand{\CCphsimb}
{
\caption{ Phase diagram from Monte Carlo simulations of the model with soft 
surface potentials in the plane of spreading pressure $\Pi$ 
(units $\epsilon/k_B \sigma_T^2$) vs. temperature $T$ (units $\epsilon/k_B$).
LE denotes disordered phase, LC-U untilted ordered phase, and LC-NN(N) 
ordered phase with tilt towards nearest or next nearest neighbours, 
respectively (undetermined).}
\label{fig:phsimb}
}


\vspace*{0.5cm}

\noindent
{\LARGE\bf
Phase Behaviour of Amphiphilic Monolayers:

\medskip

\noindent Theory and Simulation
}

\vspace{1cm}

\parbox[t]{1cm}{~}
\parbox[t]{12cm}{
D. D\"uchs, F. Schmid \\
\small
Max-Planck-Institut f\"ur Polymerforschung, Postfach 3148, 
55021 Mainz, Germany \\
Fakult\"at f\"ur Physik, Universit\"at Bielefeld, 
Universit\"atsstra\ss e 25, 33615 Bielefeld, Germany
\vspace{0.5cm}

\small
{\bf Abstract.}
Coarse grained models of monolayers of amphiphiles (Langmuir monolayers) 
have been studied theoretically and by computer simulations. We discuss
some of the insights obtained with this approach, and present new
simulation results which show that idealised models can successfully
reproduce essential aspects of the generic phase behaviour of 
Langmuir monolayers.
}

\vspace*{0.5cm}

\section{Introduction}

Amphiphilic molecules are made up of two distinct components: 
a hydrophilic part which dissolves easily in water (``loves water''), 
and a hydrophobic part which is repelled by water (``fears water''). 
In an aqueous environment, they assemble such that the hydrophobic parts 
of the molecules are shielded from the water by the hydrophilic ones. 
As a result, a rich variety of ordered and disordered structures emerge, 
featuring internal "interfaces" that separate hydrophobic from 
hydrophilic regions\cite{schick}. 

Among these, bilayer structures are receiving special attention 
because they are fundamental ingredients of biological membranes and, 
thus, basic constituents of all living organisms\cite{gennis}. 
They are typically formed by molecules which have one hydrophilic  
``head group'' attached to one or more hydrophobic hydrocarbon chains, 
e.g., lipids or fatty acids. In water, the molecules may, in certain 
parameter regions, aggregate into stacks of planar bilayers. 
These lamellar phases 
often exist in several variations: With decreasing 
temperature, they undergo a first order transition (``main transition'') 
from a high temperature ``fluid'' phase to a low temperature ``gel'' phase, 
which is characterised by higher bilayer thickness, lower chain mobility 
and higher chain ordering. Depending on the chain length and the bulkiness 
of the polar head group, different gel phases can be found, some with 
chains oriented on average perpendicular to the lamellar surface, and some with 
collectively tilted chains. In systems with bulky head groups, the strictly 
planar gel phase is often pre-empted by one with asymmetric wavy undulations 
(``ripple'' phase). Theoretical considerations have suggested that the latter 
may be related to tilt order in the bilayers\cite{lubensky}. 
From an experimental point of view, the question whether
the chains in the ripple phase are tilted or not is still debated 
\cite{sengupta,nagle}. In biological systems, membranes are usually 
maintained in the fluid state by the living organism. Nevertheless, the main 
transition is presumably of some relevance in the biological context, 
as it occurs at temperatures very close to the body temperature for some 
of the most common bilayer lipids (e.g., 41.5 ${}^0$C in DPPC). 

To assess phenomena of this type, people have been studying Langmuir 
monolayers for many years as model system that are particularly 
accessible in experiments \cite{mono_reviews}. 
Such monolayers form when amphiphilic molecules 
of sufficient chain length are spread onto an air-water interface. At low 
surface coverage, the molecules do not interact with each other and form what 
is the two-dimensional analogon of a gas. Upon compression, the system 
exhibits a first order ``gas-liquid'' transition to a phase where the 
molecules form a continuous monolayer whose behaviour resembles, in some 
sense, that of the corresponding bilayer. 
In particular, one observes a monolayer 
equivalent of the main transition -- a first order 
transition between two liquid states: the ``liquid expanded'' (LE) and the 
``liquid condensed'' (LC) state. As in the bilayer case, several phases 
are present in the condensed region, which differ in the tilt order of the 
chains, the orientational order of the backbones, and the positional 
order of the heads. In the phases which coexist with the liquid 
expanded phase, the molecules are axially symmetric and form a hexatic 
liquid. A generic phase diagram is shown in 
Figure \ref{fig:generic} \cite{generic_exp}. 

\vspace*{0.5cm}

\fig{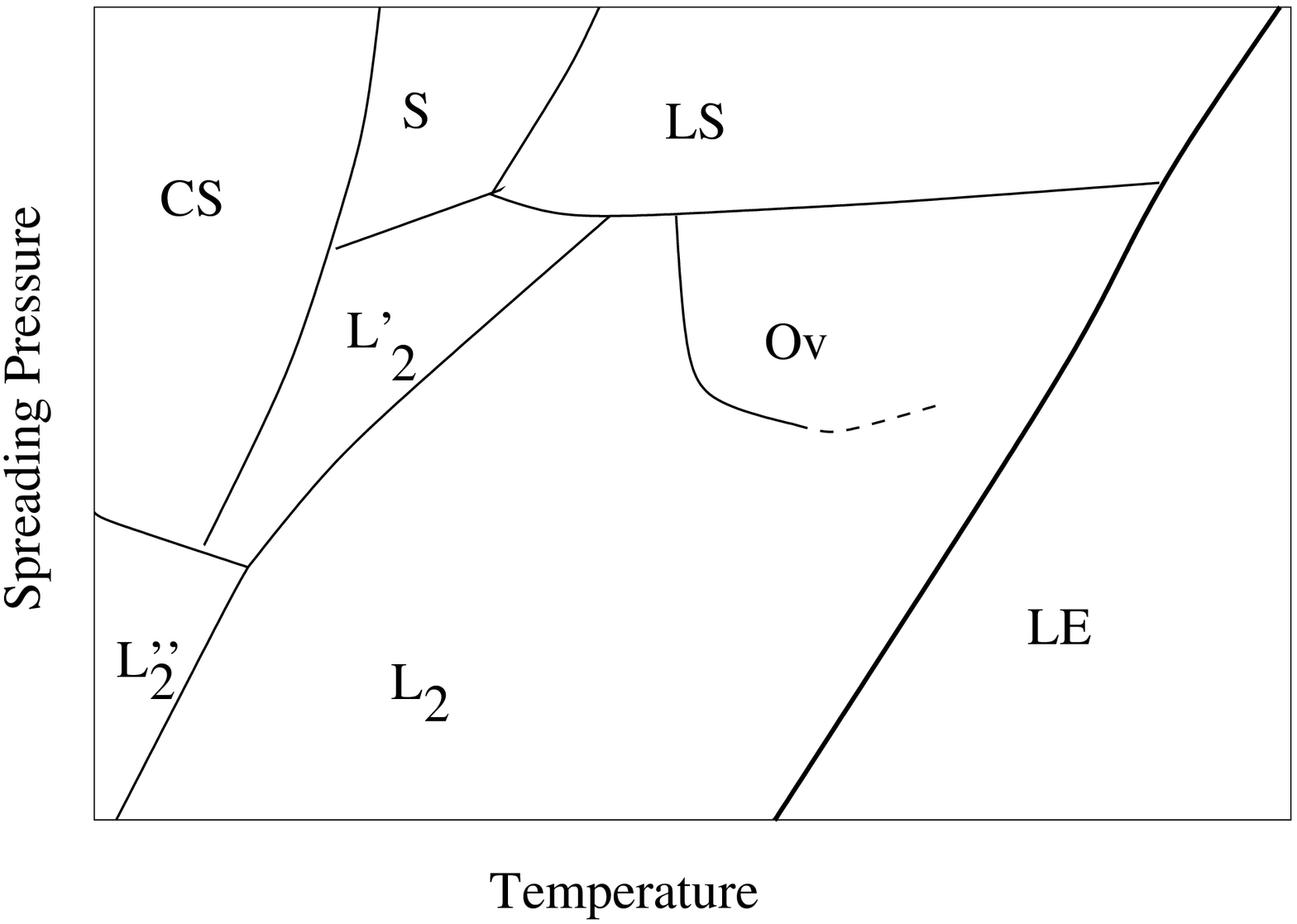}{100}{80}{}
\begin{figure}[h]
\CCgeneric
\end{figure}

Interestingly, many topological features of the phase diagram 
in the condensed region (sequence and order of phase transitions etc) 
can be understood in terms of generic Landau symmetry 
considerations\cite{generic_theory}.
The impressive results of this approach have been summarised nicely
in a recent review article by Kaganer {\em et al }\cite{kaganer}.
Here we shall focus our attention on the main transition, i.e., 
we will consider only the right part of the phase diagram with the
liquid expanded phase and the coexisting condensed phases.
Our goal is to explore possible explanations of the transitions
between those phases using simple idealised models. The paper
is organised as follows: First, we will review some theoretical findings,
then discuss recent computer simulations and present new results. 
We hope that we will convince the reader that we are now able to 
understand and reproduce the relevant characteristics of the experimental 
phase diagram quite satisfactorily.

\section{Theory}

Our theoretical work has mainly addressed the two following issues:
\begin{enumerate}
\item What is the mechanism that drives the first order fluid-fluid
transition between the liquid expanded and liquid condensed
regions?
\item Which factors determine tilt order and tilt direction?
\end{enumerate}

The first question was tackled by means of a self-consistent field
theory of a simple grafted chain model\cite{FS1,FS2}. The amphiphiles are 
modelled as stiff chains of attractive rodlike segments attached to one 
head segment, which is confined into a planar surface by a harmonic 
potential and free to move in lateral directions. This model indeed 
exhibits two coexisting liquid phases, and even an additional tilted
phase in certain parameter regions. Two ingredients are crucial in 
bringing about liquid-liquid coexistence: the flexibility of the
chains and an affinity to parallel packing (chain anisotropy). 
A typical phase diagram is shown in Figure \ref{fig:phtheory}.

\vspace*{1cm}
\fig{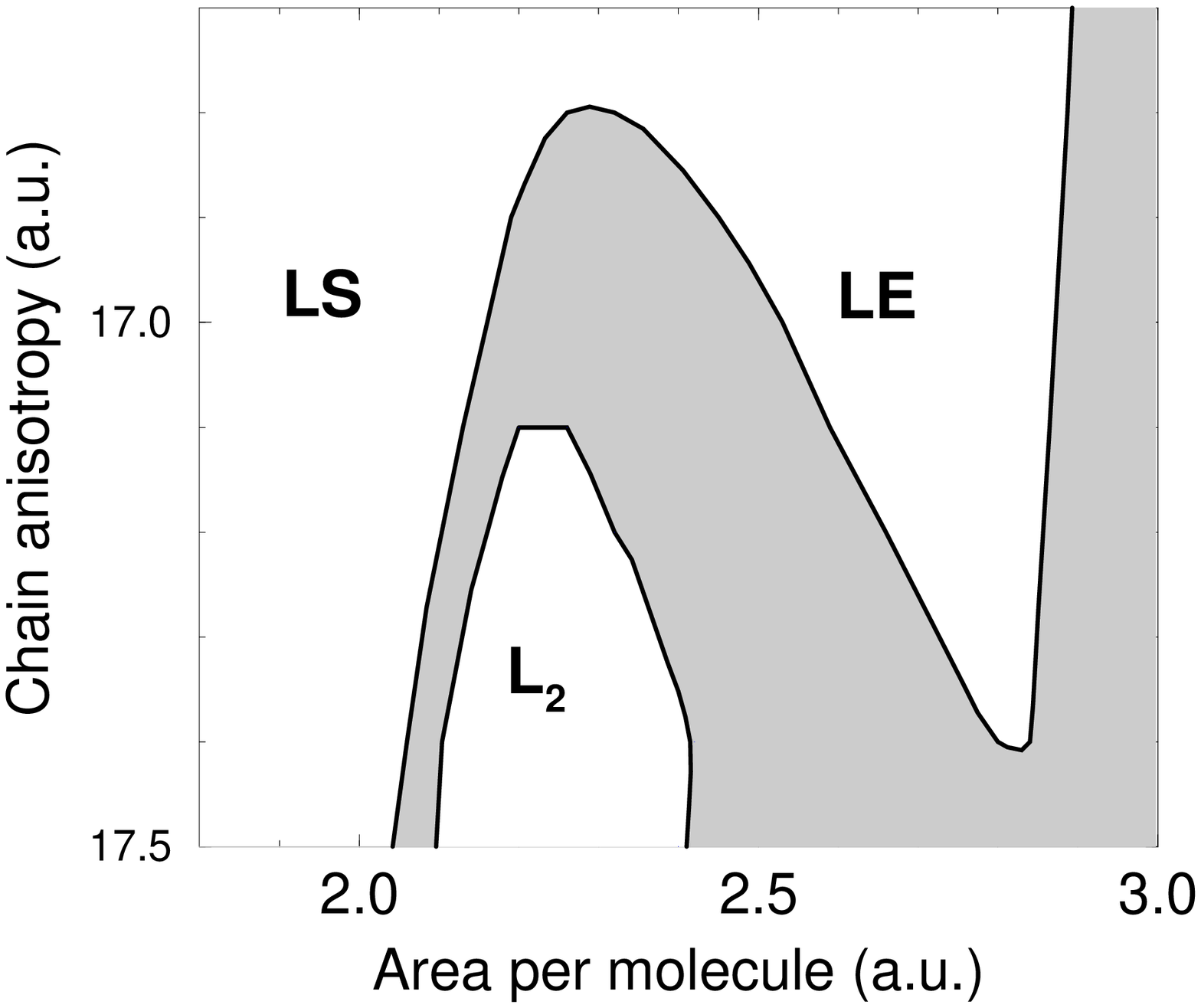}{100}{80}{}
\begin{figure}[h]
\CCphtheory
\end{figure}

Here, the phase behaviour was plotted as a function of the effective chain 
anisotropy. An almost identical phase diagram is obtained if the chain 
anisotropy is kept fixed and the chain stiffness is varied instead. 
Note that both the effective chain interactions and the effective chain 
stiffness depend on the temperature. Hence, the $y$-axis in 
Fig. \ref{fig:phtheory} can be interpreted as a temperature axis. 
Our self-consistent field calculation does not account for the 
possibility of hexatic order in the liquid condensed region. 
This is why the phase coexistence region ends in a critical point, where 
there should probably be a multicritical point followed, 
at higher temperatures, by a line of continuous 
Kosterlitz-Thouless transitions. 

Our results show that the liquid-liquid phase transition arises from 
a competition between the conformational entropy of the chains, which 
stabilises the expanded phase, and a tendency to parallel alignment, which
stabilises the condensed phase. The importance of the conformational
entropy for the transition had already been demonstrated
by experiments of Barton {\em et al} \cite{barton}: If one reduces the
chain flexibility by substituting hydrogen with fluorine,
the liquid expanded phase will disappear. 

Next, we address the issue of tilt order and tilt direction. The latter
is a well-defined quantity in a hexatic liquid since it lacks only
positional order but retains long range bond orientational 
order\cite{halperin}. We have studied tilt order in monolayers 
within an even simpler model than that sketched above, namely a system of 
rigid rods attached to head groups that are confined into a plane\cite{FS3}. 
The main effects are already apparent from an analysis of the
state of lowest energy: Tilting transitions can be induced by either  
varying the surface pressure or the head size. In both cases, 
one finds a sequence of three phases: First an untilted phase 
(small heads or high surface pressure), then a phase where the rods
are tilted towards next-nearest neighbours, and finally (large heads
and low surface pressure) a phase with tilt towards nearest neighbours. 

That precise sequence is found experimentally in the pressure-temperature
phase diagram (see Figure \ref{fig:generic}). The argument predicts
that the phase with tilt towards nearest neighbours should be suppressed
if the head groups are too small. This has indeed been observed in
experiments, where the effective head size was reduced by increasing
the pH of the subphase\cite{shih}, or by replacing the COOH head
groups of fatty acids by smaller alcohol head groups\cite{teer1,teer2}.

Note that the theoretical predictions were obtained using a simple model 
of cylindrical rigid rods. A much more complex ground-state phase behaviour 
results if in addition the rods are given internal structure. For a model 
that uses rigid beaded rods, Opps et al have found a diversified occurrence
of NN to NNN phases depending on the head/tail diameter ratio, bond lengths, 
and interaction potentials: The head diameters govern the overall tilting 
behaviour, whereas the finer details of the phase diagram depend on the 
precise nature of the interaction potentials\cite{opps}. 

To summarise this section, our theoretical studies have shown
that much of the phase behaviour in Langmuir monolayers can be discussed
in terms of a few elementary properties of amphiphiles: The flexible
chains with their tendency to parallel packing drive the 
transition from liquid expanded to liquid condensed, and the tilting 
transitions are driven by an interplay 
between head size, chain diameter, and surface pressure. 

In the next section, we will discuss computer simulations of a model
which incorporates just these few basic ingredients.

\section{Computer simulations}

A vast amount of activity has been devoted to the simulation of 
surfactant systems in general, and bilayers or monolayers in particular.
For a general account, we refer the inclined reader to recent reviews 
\cite{sim1,sim2,sim3} and shall only report on our own work here
\cite{christoph1,christoph2,dominik1}.

We model the amphiphiles as chains of $N$ beads with diameter $\sigma_T$, 
attached to one slightly larger head bead with diameter 
$\sigma_H$, which is confined to the plane $z=0$. Beads are not allowed into 
the half space $z<0$. Two beads that are not direct neighbours in the same 
chain interact with truncated and shifted Lennard-Jones potentials
\begin{equation}
\label{vlj}
V_{LJ}(r) = \left\{ \begin{array}{lcr}
\epsilon \cdot
\Big( \big(\sigma/r\big)^{12} - 2 \big(\sigma/r\big)^6 + v_c \Big) &
\mbox{for} & r \le R_0 \\
0 & \mbox{for} & r > R_0
\end{array} \right. ,
\end{equation}
where the offset $v_c$ is chosen such that $V_{LJ}(r)$ is continuous
at $r=R_0$, and the cutoff $R_0$ is $R_0 = 2 \sigma_T$ for the tail beads and
$R_0 = \sigma_H$ for the head beads. Hence, tail beads attract each other
and head beads are purely repulsive. The interactions between head and tail 
beads are repulsive with the effective diameter $(\sigma_T + \sigma_H)/2$.
Beads are connected by springs of length $d$ subject to the weakly nonlinear
spring potential
\begin{equation}
\label{fene}
V_{S}(d) = \left\{ \begin{array}{l c r}
- \frac{k_{S}}{2} \; d_{S}^2 \; \ln\Big[ 1 - (d-d_0)^2/d_{S}{}^2 \Big]
& \mbox{for} & |d-d_0|<d_{S} \\
\infty & \mbox{for} & |d-d_0| > d_{S}
\end{array} \right.
\end{equation}
Moreover, a stiffness potential 
\begin{equation}
\label{ba}
V_{A} = k_{A} \cdot (1- \cos \theta)
\end{equation}
is imposed, which acts on the angle $\theta$ between subsequent springs 
and favors $\theta=0$ (straight chains). Unless stated otherwise, the model 
parameters are $d_0 = 0.7 \sigma_T$ (equilibrium spring length), 
$d_S = 0.2 \sigma_T$, $k_S = 100 \epsilon$, $k_A = 10 \epsilon$, and 
$\sigma_H = 1.1 \sigma_T$. In most cases, systems of 144 chains with a total 
length of 7 beads were studied. A preliminary discussion of chain end 
and system size effects can be found in Ref. \cite{christoph2}. The 
simulations were conducted at constant spreading pressure $\Pi$ in a 
simulation box of variable size and shape, with periodic boundary conditions 
in the $x$ and $y$ directions.

Quantities of special interest are the collective tilt of the chains 
and the liquid structure. The collective tilt is measured with the 
order parameter
\begin{equation}
R_{xy} = \sqrt{\langle [x]^2 + [y]^2 \rangle },
\end{equation}
where $[x]$ and $[y]$ denote the $x$ and $y$ components, respectively, 
of the head-to-end vector of a chain, averaged over all chains in one 
configuration, and $\langle \cdot \rangle$, the statistical average
over all configurations. To study the liquid structure, we have inspected
radial pair correlation functions and the hexagonal order parameter
of two-dimensional melting
\begin{equation}
\label{psi}
\Psi_6 = \bigg\langle \bigg| \frac{1}{6n}
\sum_{j=1}^n \sum_{k=1}^6 \exp(i 6 \phi_{jk}) \bigg|^2 \Big\rangle,
\end{equation}
which measures the orientational long-range order of nearest-neighbour
directions. Here the sum $j$ runs over all heads of the system, the sum $k$,  
over the six nearest neighbors of $j$, and $\Phi_{jk}$ is the angle
between the vector connecting the two heads and an arbitrary
reference axis. 

At head size $\sigma_H = 1.1 \sigma_T$, we find four different 
phases: a disordered liquid (LE) and three condensed phases,
one without tilt (LC-U), one with tilt towards nearest neighbors (LC-NN)
and one with tilt towards next nearest neighbors (LC-NNN). 
Our systems are too small to allow for dislocations, and the molecules
are almost always arranged on a defect free lattice in the condensed
region. However, hexatic disorder may well be present in larger systems. 
The phase diagram in the pressure-area plane, obtained by inspection of 
the order parameters $R_{xy}$ and $\Psi_6$, as well as by a phonon expansion
at low temperatures\cite{christoph1}, is plotted in 
Figure \ref{fig:phsima}. 

\vspace*{4cm}

\fig{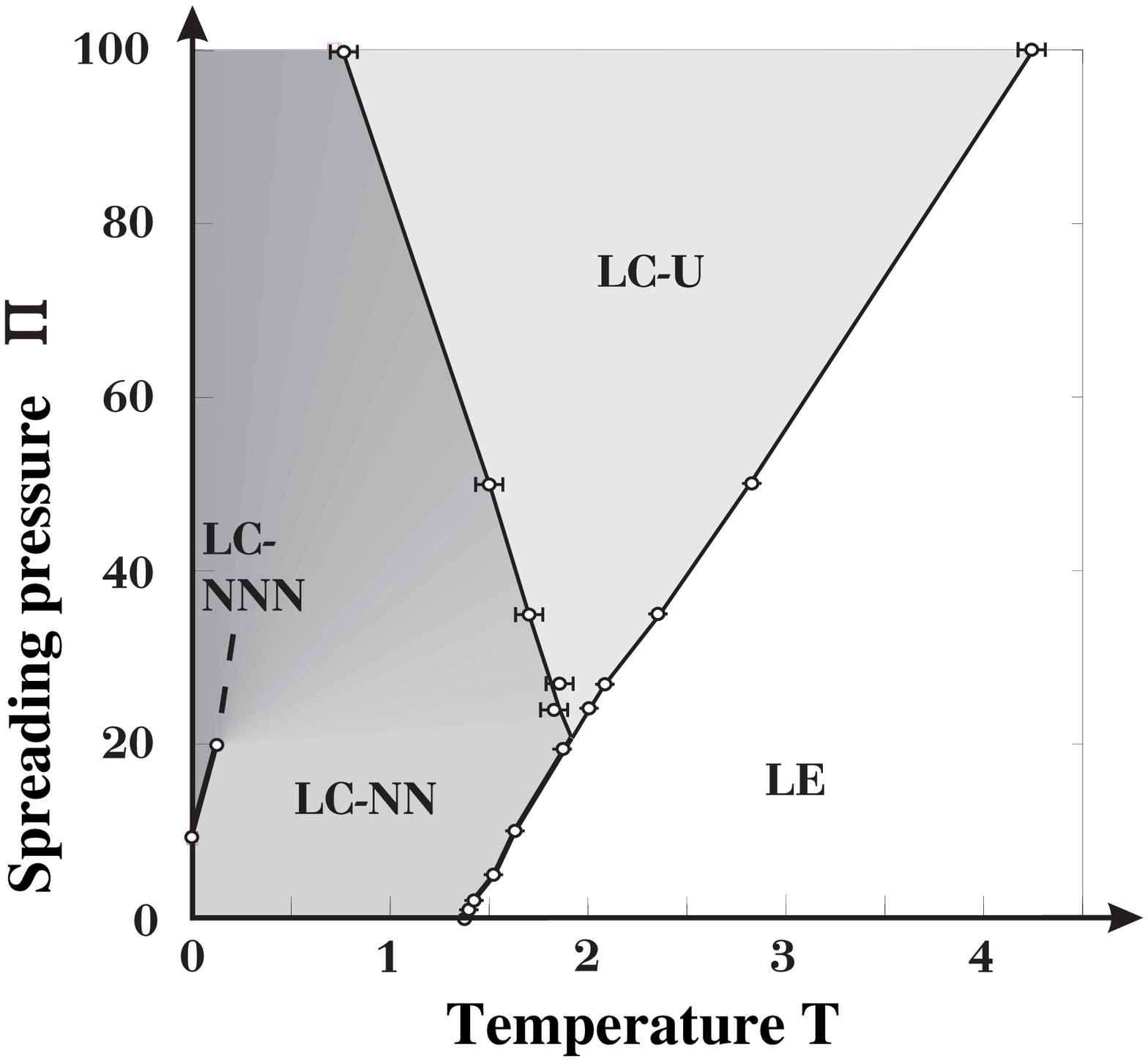}{100}{80}{}
\vspace*{-2cm}
\begin{figure}[h]
\CCphsima
\end{figure}

We are now in a position to compare the simulations with some of 
the theoretical results. Figure \ref{fig:rpvf} shows examples of radial 
pair correlation functions for head beads and whole chains at temperatures 
well below, slightly below, slightly above, and well above the LE/LC transition 
temperature. One notices that the correlation function of whole chains
changes quite dramatically at the phase transition, whereas the head 
correlation function remains rather unaffected. The chains maintain
the order below the transition, and promote disorder above the transition.
In agreement with the theoretical prediction, 
one can conclude that the chains drive the transition.

\fig{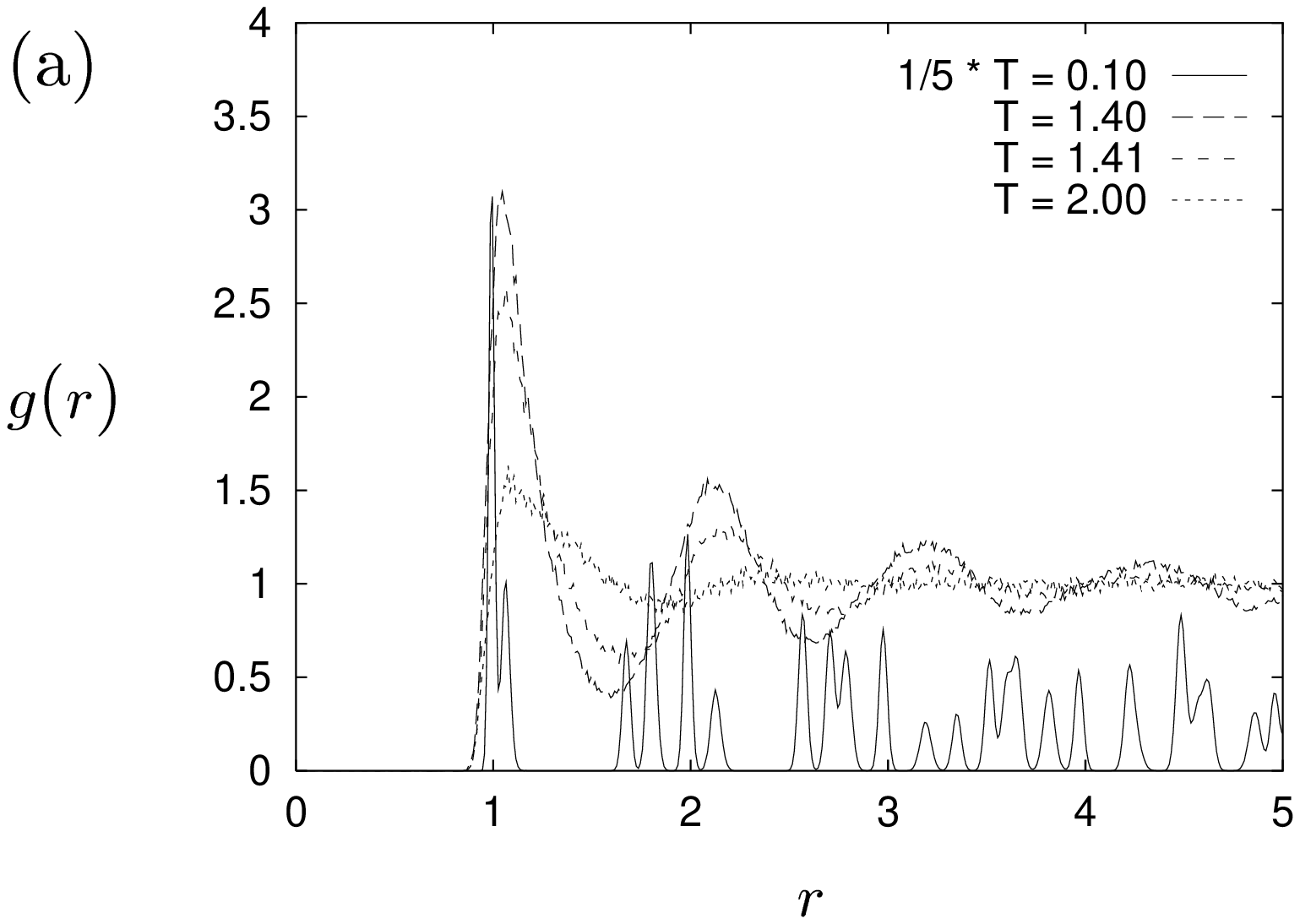}{120}{80}{}
\vspace*{-2cm}
\fig{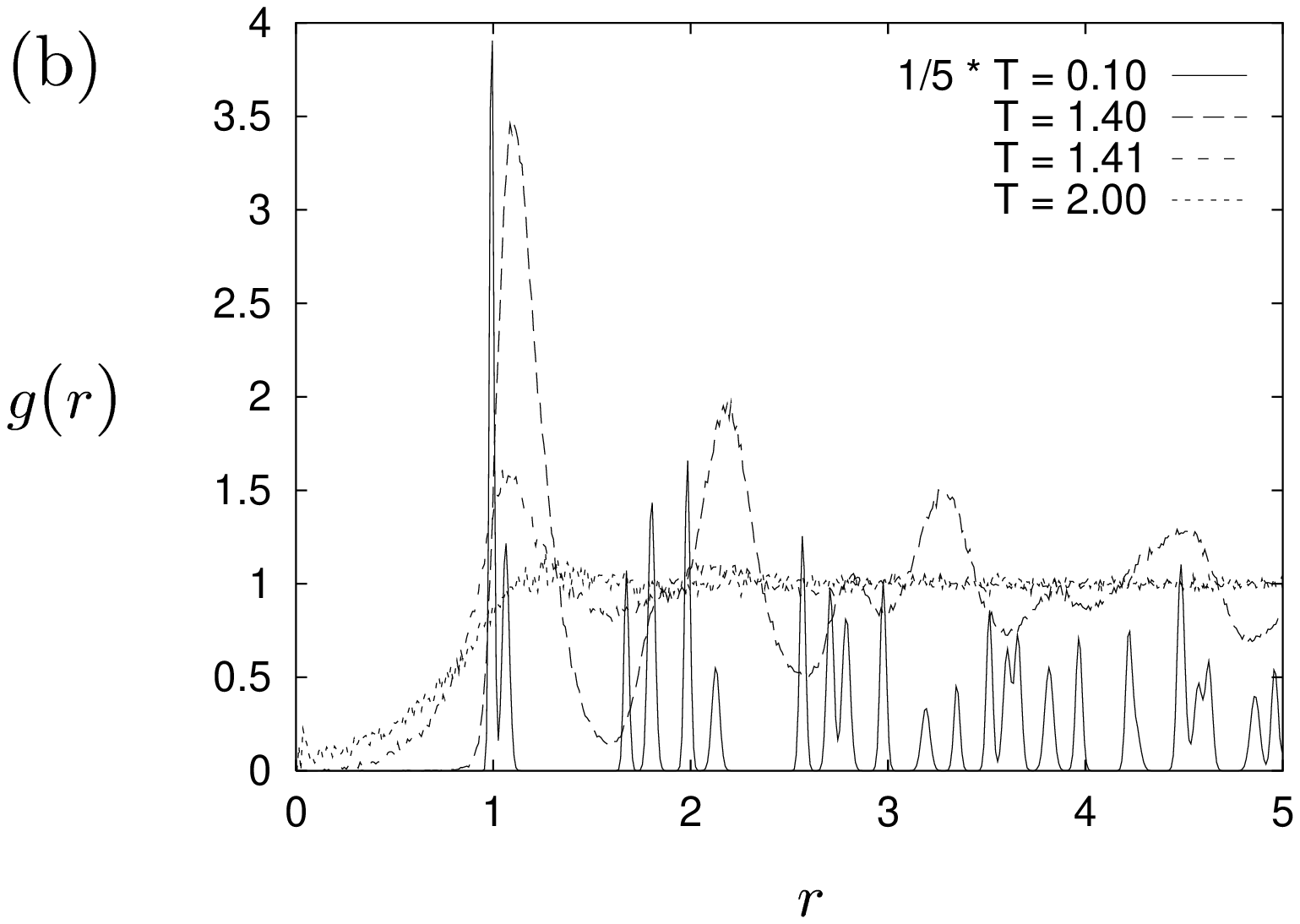}{120}{80}{}
\vspace*{-1cm}
\begin{figure}[h]
\CCrpvf
\end{figure}

At low temperatures, the model exhibits the sequence of tilting transitions 
predicted by the theory (see Figure \ref{fig:phsima}),
with a first order transition between the two tilted phases.
At higher temperatures, the situation is less clear. The direct inspection 
of several configuration snapshots suggests that the system might pass 
directly from the LC-NN phase to the LC-U phase, skipping the intermediate
LC-NNN state. Unfortunately, the tilt direction fluctuates so strongly that 
the average direction cannot be determined unambiguously. Simulations of much 
larger systems would be needed to clarify this aspect of the phase diagram.

If one increases the head size, the region where the chains tilt towards 
nearest neighbours becomes larger, as anticipated by the theory. Interestingly, 
this goes along with the appearance of a new, unexpected LC-NN modification: 
a modulated striped phase\cite{christoph2}. It proves to be extremely 
stable over a wide parameter region. 
One may speculate on their existence in real systems.

Although the phase diagram of Figure \ref{fig:phsima} is already 
gratifyingly similar to the experimental phase diagram 
(Figure \ref{fig:generic}), it still contains one obvious flaw: The pressure 
at the transition between tilted and untilted phases is largely temperature
independent in experiments, whereas it has a considerable slope in the 
simulations. The most plausible explanation for this discrepancy is to 
attribute it to the overly simple treatment of the head groups, or, 
more precisely, to the rigid constraints imposed on them. The 
slope of the phase boundary can easily be rationalised if one assumes
that the heads are forced to absorb most of the pressure because 
they cannot move out of their plane.

In order to remedy this situation, we have conducted a set of simulations
where the surface constraints are softened up and replaced by harmonic 
surface potentials\cite{dominik1}. The main results of this study 
shall be presented now.

The new surface potentials were chosen as follows. Head beads are
subject to a potential
\begin{equation}
\label{vh}
V_{h}(r) = \left\{ \begin{array}{lcr}
0 & \mbox{for} & z < -0.5 W\\
-{\epsilon_h}/{2} \ln (1-(z+0.5 W)^2/W^2)
& \mbox{for} & -0.5 W < z < 0.5 W
\end{array} \right. ,
\end{equation}
and tail beads to a potential
\begin{equation}
\label{vt}
V_{t}(r) = \left\{ \begin{array}{lcr}
-{\epsilon_t}/{2} \ln (1-(z-0.5W)^2/W^2)
& \mbox{for} & -0.5 W < z < 0.5 W \\
0 & \mbox{for} & z > 0.5 W\\
\end{array} \right. .
\end{equation}
The width $W$ of the potential is set to 1 $\sigma_T$, and
the strength factors $\epsilon_h$ and $\epsilon_t$ are
given the values $10 \epsilon$. As we will see, the exact form of 
the surface potentials is not essential.

Apart from this innovation, the model is defined as before, with the one 
exception that the stiffness potential was reduced to $k_A = 4.7 \epsilon$.
This is the value which one would estimate from the Rigby-Roe model for 
hydrocarbon chains\cite{roe}, assuming that two carbon atoms correspond 
roughly to one bead in our model. The size of the head bead was chosen as 
$\sigma_H = 1.1 \sigma_T$, as in the study discussed above.

The results can be summarised as follows: 

We find essentially the
same phases and the same phase characteristics as before. As an
unwanted artefact of the model, one observes at low temperatures and high 
pressures a double peak structure in the head density profile $\rho_h(z)$.
Fortunately, the effect disappears at temperatures 
$T > 0.5 \epsilon/k_B$, and the system is well behaved at all 
parameter ranges of interest. 

As we had hoped, the transition pressures of the tilting transition
at lower temperatures drop considerably. In order to explore the sensitivity 
of the phase behaviour to the parameters of the new surface potentials, we 
have performed a few simulation runs of systems with double potential
width $W$. Results for an exemplary isotherm are shown in 
Figure \ref{fig:rxy_2}: The phase transition occurs at almost
the same pressure in systems with potential width $W = \sigma_T$ or 
$W = 2 \sigma_T$. The transition pressure is much lower than
that in the original model (cf. Figure \ref{fig:phsima}).
Hence, a dramatic lowering of the transition pressures 
is achieved by a mere relaxation of the head groups. 
Once this lowering is accomplished, further relaxation does not have 
much impact.

\fig{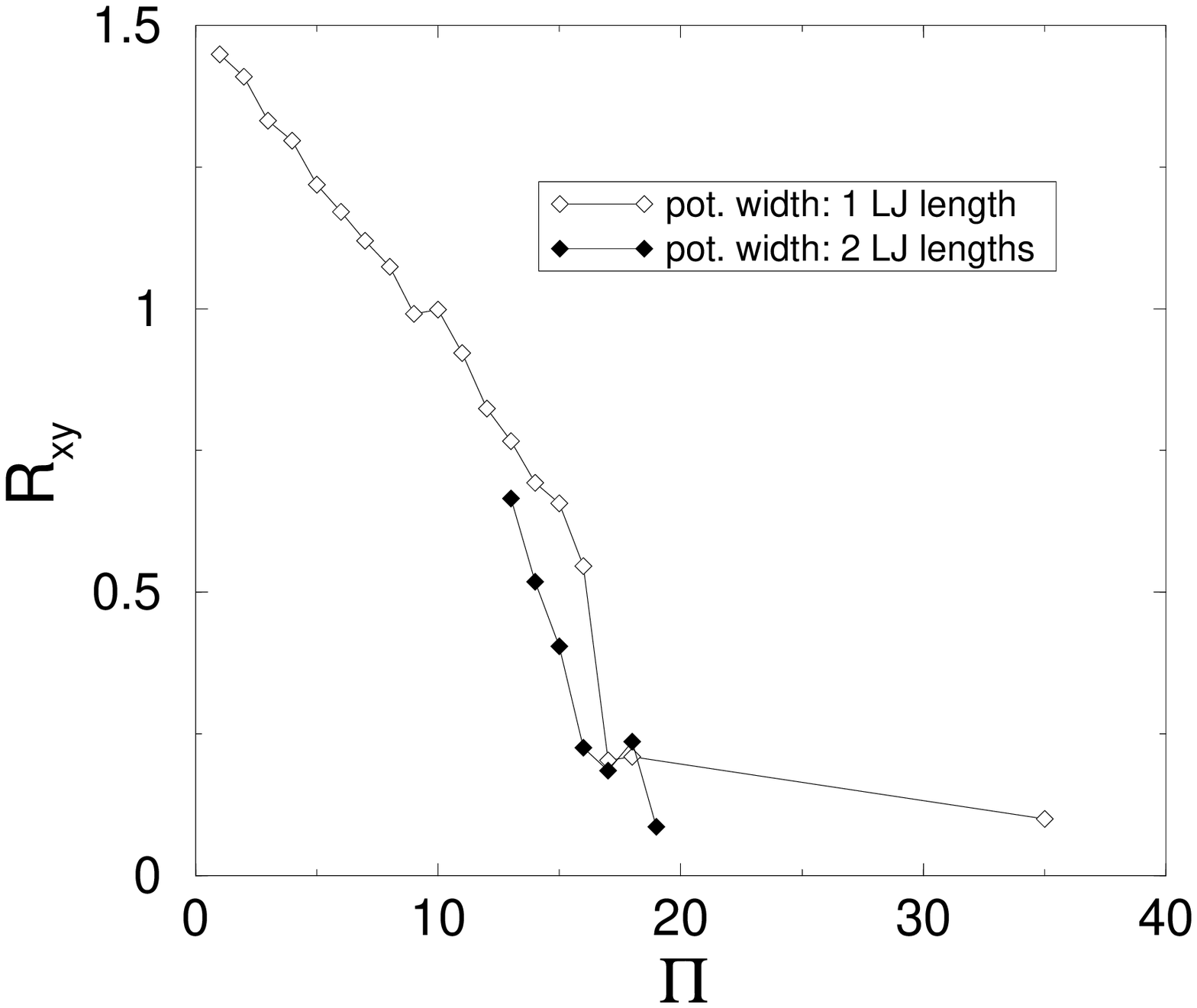}{80}{70}{}
\vspace*{-1cm}
\begin{figure}[h]
\CCrxy
\end{figure}

Another new feature of the model, compared with the earlier version, 
concerns the order of the tilting transition. Whereas in the old 
version, we had no reason to doubt that it is continuous (although
this really ought to be established rigorously by a finite size analysis, 
of course), the order parameter $R_{xy}$ now seems to jump between two
states in the vicinity of the transition. A typical histogram of $R_{xy}$ 
is shown in Figure \ref{fig:rxy_hist}. One clearly discerns two peaks, 
corresponding to two states of different tilt order. This observation
suggests that the tilting transition might be first order. 
Again, simulations of much larger systems would be needed to 
corroborate this suspicion.

\fig{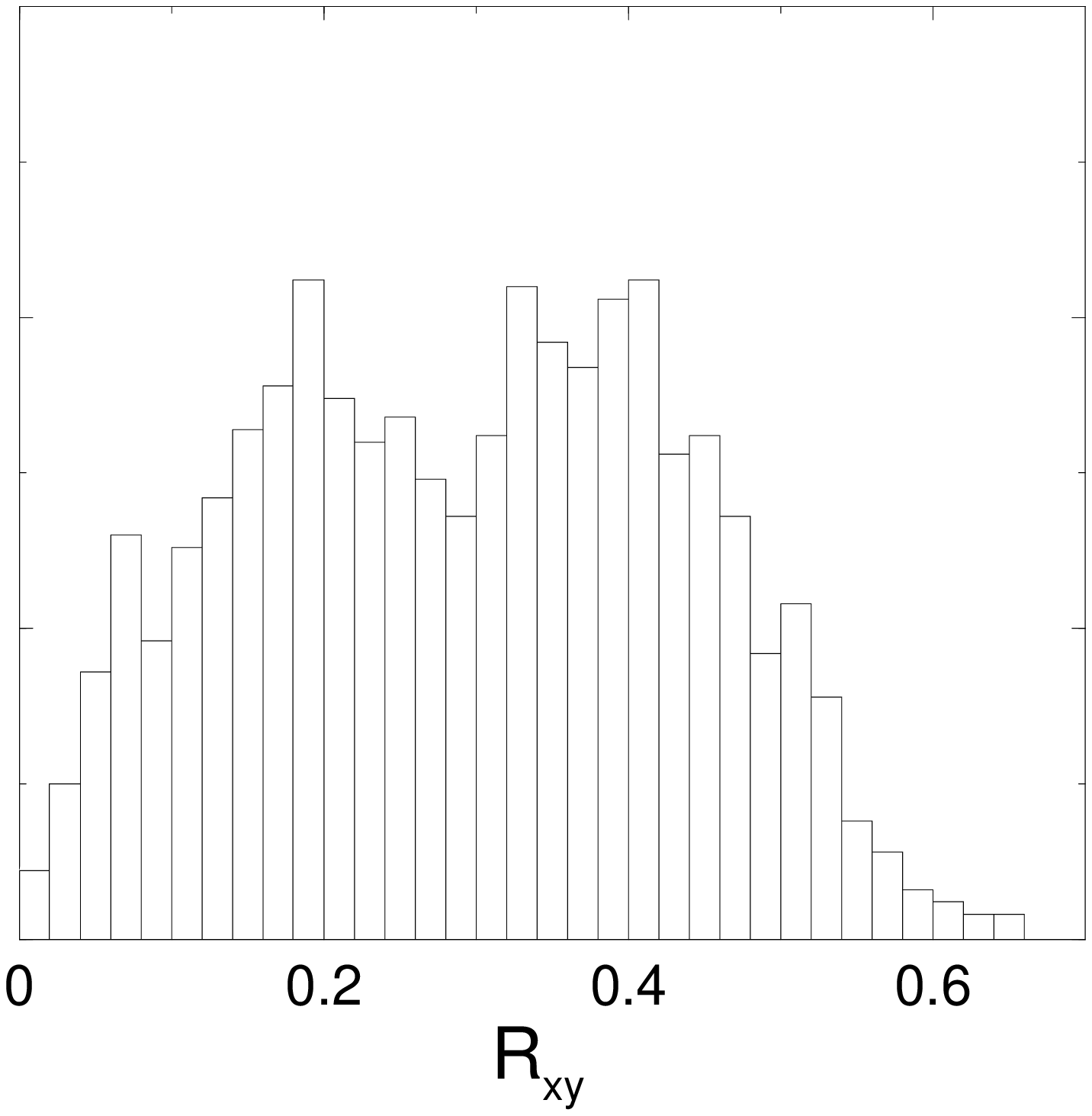}{80}{75}{}
\vspace*{-1.5cm}
\begin{figure}[h]
\CCrxyh
\end{figure}
\clearpage

The phase diagram of the revised model is shown in Figure \ref{fig:phsimb}. 
The changes to the LE/LC boundary are rather marginal compared to
the earlier version (Figure \ref{fig:phsima}); it experiences
only a small shift to lower temperatures. However, the boundary between 
the tilted and untilted phases is affected in the desired dramatic
way: The transition pressures at lower temperatures drop considerably,
and the slope of the transition line is now almost flat, like in
experiments. 

\fig{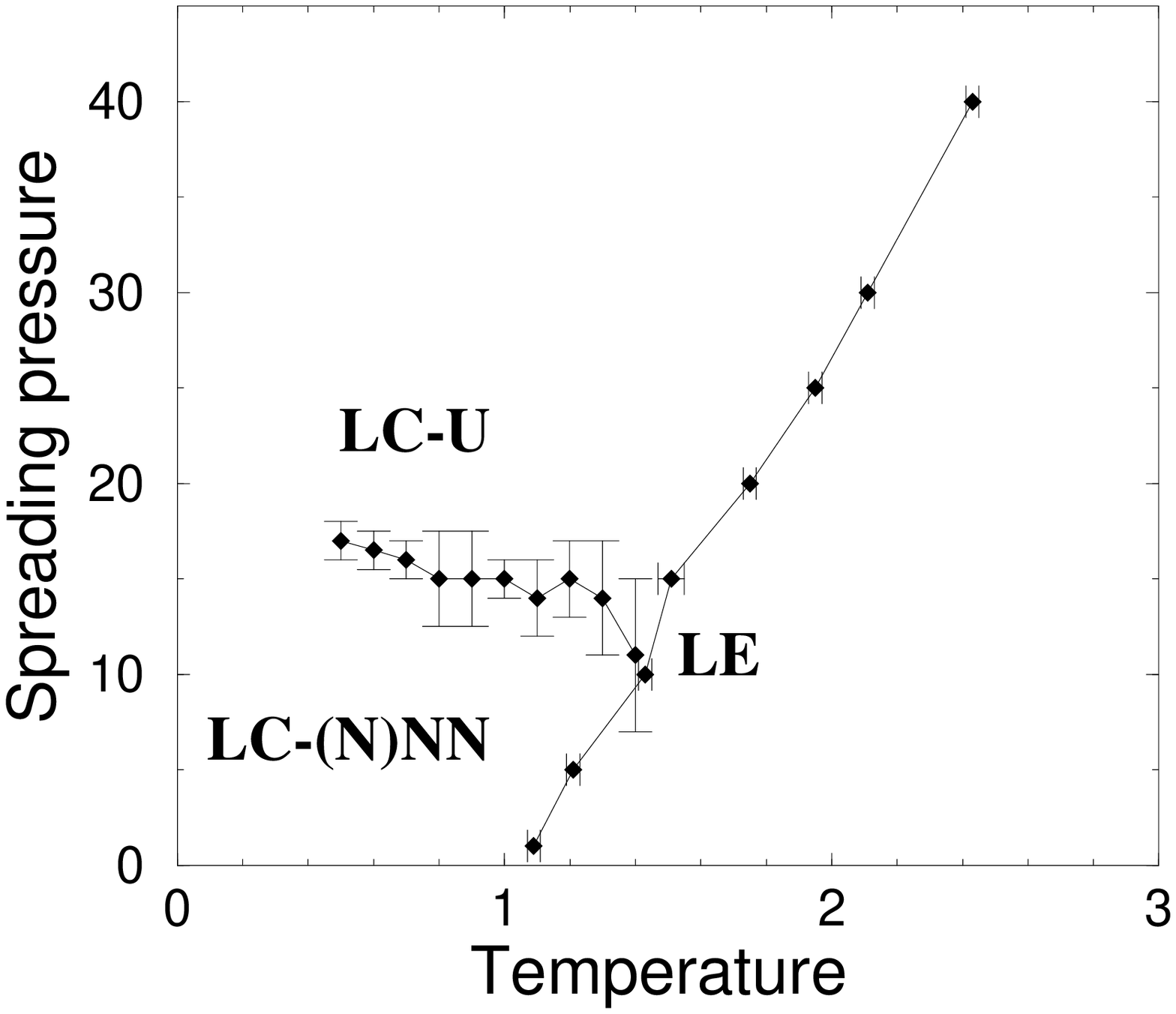}{80}{80}{}
\begin{figure}[h]
\CCphsimb
\end{figure}

We conclude that we have established a minimal model which reproduces 
in computer simulations the essential features of the experimentally 
observed generic phase behaviour of Langmuir monolayers. Details of the
phase diagram will still need to be established by systematic finite size 
studies. We note that long-range Coulomb and dipolar interactions are
not included in the model so far. In the coexistence region of the liquid
expanded and liquid condensed states, the interplay between electrostatic
interactions and line tensions leads to a variety of interesting domain 
patterns on a mesoscopic scale\cite{mono_reviews}. On the microscopic
scale considered here, however, these long-range interactions seem 
less influential. We feel that the general agreement between the phase 
behaviour of the model and the experimental one is now satisfactory 
enough for us to use the model as a basis for the investigation of more 
complex problems.

\section*{Acknowledgements}

Part of the work presented here was done in collaboration with
M. Schick, C. Stadler, and H. Lange. We thank N. Akino, K. Binder, 
F. M. Haas, R. Hilfer, and S. Opps for enjoyable interactions
and useful discussions.

\end{document}